\documentclass[10pt]{article}
\usepackage{hyperref}
\usepackage{graphicx}
\usepackage{graphics}
\usepackage{dsfont}
\usepackage{epsfig}
\usepackage{bbm}
\usepackage{amsmath,amssymb,amsthm,amscd}
\usepackage[sort&compress,square,numbers]{natbib} 
\usepackage{float}
\usepackage{braket}

\usepackage[OT2,T1]{fontenc}
\DeclareSymbolFont{cyrletters}{OT2}{wncyr}{m}{n}
\DeclareMathSymbol{\Sha}{\mathalpha}{cyrletters}{"58}

\restylefloat{table}

\newcommand{\nn}{{\nonumber}}

\def\Mgn[#1]#2{{\overline{\cal M}_{#1,#2}}}
\def\pqs[#1,#2]{{\footnotesize{$\left[\begin{array}{c} #1\\#2  \end{array}\right]$}}} 
\def\pqsu[#1,#2]{\left[\begin{array}{c} #1\\#2  \end{array}\right]} 
\def\pqssu[#1,#2]{{\footnotesize{\left[\begin{array}{c} #1\\#2  \end{array}\right]}}} 
\def\pqh[#1,#2]{{\footnotesize{$\left[\begin{array}{c} #1\\#2  \end{array}\right]$}}} 
\def\pqhu[#1,#2]{\left[\begin{array}{c} #1\\#2  \end{array}\right]}

\newcommand{\ba}{\begin{eqnarray*}}
\newcommand{\ea}{\end{eqnarray*}}
\newcommand{\ban}{\begin{eqnarray}}
\newcommand{\ean}{\end{eqnarray}}
\newcommand{\be}{\begin{equation}}
\newcommand{\ee}{\end{equation}}
\newcommand{\ben}{\begin{equation}}
\newcommand{\een}{\end{equation}}
\numberwithin{equation}{section}

\numberwithin{equation}{section}

\begin{document}
\rightline{
CERN-PH-TH-2015-034}
\rightline{
UPR-1270-T}
\vskip 1 cm
\centerline{\Large \bf F-Theory Vacua with $\mathbb{Z}_3$ Gauge Symmetry}  \vskip 0.5 cm

\vskip 10pt \centerline{ 
\textbf{Mirjam Cveti\v c\footnote{cvetic@cvetic.hep.upenn.edu}$^{a}$, Ron Donagi\footnote{donagi@math.upenn.edu}$^{a,b}$,  Denis Klevers\footnote{Denis.Klevers@cern.ch}$^{c}$, }} \vskip .1cm
\centerline{
\textbf{Hernan Piragua\footnote{hpiragua@sas.upenn.edu}$^{a}$, Maximilian Poretschkin\footnote{mporet@sas.upenn.edu}$^{a}$}} 

\begin{center}
\small
{$^{a}$ \textit{Department of Physics and Astronomy, University of Pennsylvania,}}\\ 
{\textit{Philadelphia, PA 19104-6396, USA}}\\ [3 mm]
{$^{b}$ \textit{Department of Mathematics, University of Pennsylvania,}}\\ 
{\textit{Philadelphia, PA 19104-6396, USA}}\\ [3 mm]
{$^{c}$ \textit{Theory Group, Physics Department, CERN, CH-1211, Geneva 23, Switzerland}}\\ 
\end{center}

\begin{abstract}
Discrete gauge groups naturally arise in F-theory compactifications on genus-one fibered Calabi-Yau 
manifolds. Such geometries appear in families that are parameterized by the Tate-Shafarevich 
group of the genus-one fibration. While the F-theory compactification on any element of this family gives rise to the same physics, the corresponding M-theory compactifications on these geometries differ and are obtained by a fluxed circle reduction of the former. In this note, we focus on an element of order three in the Tate-Shafarevich group of the general cubic.  We discuss how the different M-theory vacua and the associated discrete gauge groups can be obtained by Higgsing of a pair of five-dimensional U(1) symmetries. The Higgs fields arise from vanishing cycles in $I_2$-fibers that appear at certain codimension two loci in the base. We explicitly identify all three curves that give rise to the corresponding Higgs fields. In this analysis the investigation of different resolved phases of the underlying geometry plays a crucial r\^ole.
\end{abstract}

\section{Introduction}
Discrete symmetries play a key r\^ole
for constructing extensions of the standard model of particle physics. 
In particular, 
discrete symmetries  are used to forbid terms in the MSSM superpotential that would allow for fast 
proton decay or other processes which are highly suppressed in the standard model. Well known 
examples are provided by R-parity ($\mathbb{Z}_2$), baryon triality ($\mathbb{Z}_3$) and 
proton hexality ($\mathbb{Z}_6$) \cite{Ibanez:1991hv, Ibanez:1991pr,Dreiner:2005rd}. Conceptually, these discrete symmetries must be realized as 
discrete gauge symmetries arising, for example, as remnants of a broken U(1) symmetries because 
global discrete symmetries cannot exist in a theory of quantum gravity \cite{Banks:2010zn}. 

The understanding of the geometrical origin of discrete gauge symmetries in F-theory compactifications 
is therefore of crucial interest both for conceptual as well as for phenomenological reasons. Here, discrete 
gauge symmetries arise from Calabi-Yau geometries which are only genus-one fibrations without section, 
in contrast to elliptic fibrations with sections.  Recently, there has been progress in understanding the 
physics of such compactifications, starting with \cite{Braun:2014oya} and followed by 
\cite{Morrison:2014era, Anderson:2014yva, Garcia-Etxebarria:2014qua, Braun:2014qka, Mayrhofer:2014haa, Mayrhofer:2014laa, Klevers:2014bqa}. A natural object which is attached to this kind of 
compactifications is given by the Tate-Shafarevich (TS) group of the genus-one fibration which is a 
discrete group that organizes inequivalent genus-one geometries which share the same associated Jacobian 
fibration. An F-theory compactification on a genus-one fibration does only depend on the
Weierstrass equation, that is the Jacobian fibrations, and is, thus, insensitive to the chosen element of 
the TS-group. The TS-group  determines the resulting discrete gauge symmetries of the F-theory 
compactification.  In contrast, M-theory compactifications on different elements in the TS-group of the 
genus-one fibration are physically distinct. This is consistent with M-/F-theory duality due to the additional degree of freedom of  
the Wilson line of the discrete group in the circle compactification  from F-theory to  M-theory 
\cite{de Boer:2001px}.

Genus-one fibrations do not posses a section, but only multi-sections.
A multi-section defining an $n$-fold branched cover over the base of the genus-one fibration is referred 
to as a $n$-section. The moduli space of F-theory compactifications on genus-one fibrations with an 
$n$-section is connected by an extremal transition to that of compactifications on elliptic fibrations 
with $n$ sections. Explicit models with up to three U(1) factors have been systematically constructed 
and analyzed for F-theory in \cite{Borchmann:2013jwa,Cvetic:2013nia,Cvetic:2013jta,Cvetic:2013qsa}. 
This allows an analysis of the physics using the two dual pictures before and after the 
transition. This transition has been identified in the analyzed situations and for Calabi-Yau threefolds
as a conifold transition \cite{Anderson:2014yva, Garcia-Etxebarria:2014qua}. 
The vanishing cycles in this transition are rational curves that appear as 
components of $I_2$-fibers which occur at certain codimension two loci in the base of the elliptic 
fibration.  The resulting singular geometry is deformed by a three-sphere 
$S^3$ which glues several sections into a multi-section \cite{Morrison:2014era}. From the effective 
field theory point of view, in the blow-down appear a number of massless fields from
M2-branes wrapping the vanishing cycle, that acquire a vacuum expectation value (VEV) corresponding 
to the deformation. 
In particular, the Higgsing of a U(1)$^n$ symmetry to $\mathbb{Z}_n$ symmetry requires for the final 
step in the chain of Higgsings with a single remaining U(1) a massless scalar field of charge $n$. 
When compactifying on a circle to M-theory, there is a choice of Wilson line for this U(1). 
There are $n$ different choices to turn on a Wilson line along the compactification 
circle yielding $n$ different vacua of M-theory. Indeed, as this Wilson line also affects the mass formula 
for the Kaluza-Klein modes one can see that there will be  $n$ inequivalent Kaluza Klein towers that can 
arise in five dimensions. Notably for each choice of Wilson line, there is one Kaluza-Klein mode within 
each tower that becomes massless and serves as five-dimensional Higgs field which - once they acquire 
a vacuum expectation value - lead to $n$ different five-dimensional vacua. These are to be compared 
with the $n$ different M-theory compactifications on the respective elements of the TS group.
Geometrically, one thus has to identify $n$ different vanishing curves that correspond to these
$n$ Kaluze Klein modes of the Higgs field.

Most of the recent works have focused on the case of the gauge group $\mathbb{Z}_2$ which is the
simplest case to consider as the TS-group consists in this case only of the genus one 
fibration (realized as a hypersurface in $\mathbb{P}[1,1,2]$) and its Jacobian. Accordingly, it is 
possible to identify the two corresponding Higgs fields as arising from the two rational components of 
the same $I_2$-fiber \cite{Mayrhofer:2014haa, Mayrhofer:2014laa}. 

In this note we extend the above analysis beyond $\mathbb{Z}_2$ concentrating mainly on the case of 
$\mathbb{Z}_3$ which arises from the most general cubic admitting a trisection only, although we 
expect our findings will also have to play a key role for the understanding of  
TS-groups of higher order.
In the case of $\mathbb{Z}_3$, there are three different M-theory vacua. 
We pursue the strategy to geometrically identify three different curves that give rise to the three 
different 5D Higgs fields in M-theory that lead to these inequivalent vacua. The main obstacle in 
identifying  these curves is the fact all three curves should be located in the reducible  $I_2$-fiber at 
the same codimension two locus in the two-dimensional base. However, an $I_2$-fiber naively has only 
two rational curves, so that the presence of the wanted third curve is elusive. 
The key result of this note is to identify the third rational curve inside the $I_2$-fiber.

We will use two different methods to demonstrate its existence. As a first evidence, we compute
a non-zero Gromov-Witten invariant for the elliptic fibration in the class of the purported third curve, 
that, as expected, agrees with the number of the relevant $I_2$-loci. As a stronger check, we consider 
two different phases of the resolution of the elliptic fibration before the Higgsing. The corresponding geometry is the $dP_1$-elliptic fibration considered in \cite{Klevers:2014bqa}. In other words, we 
extend the above conifold picture by also taking flop transitions\footnote{It is worth mentioning that we have to consider a non-toric flop.} into account\footnote{A similar phenomenon has 
already been observed in \cite{Witten:1996qb} in the context of tensionless strings in six and five 
dimensions. In that case too, M-theory is able to see the different phases of the theory arising from flop 
transitions, while these coincide in the six-dimensional F-theory compactification.}.
In one resolution phase, we are able to manifestly see two  curves of the $I_2$-fiber leading to two 
inequivalent vacua after the transition, while the third curve remains invisible. In the second resolution 
phase, we see again two curves in the $I_2$-fiber, however, one of which was not present in the first 
phase and leading to the \textit{third} 
vacuum, while one of the original two curves in the first resolution has become invisible.   
In addition, we argue that there are also in the case of TS-group $\mathbb{Z}_3$
just two different geometries, namely the most general cubic and its Jacobian. However, the former can 
be equipped with two different actions of the Jacobian which 
make them differ as elements of the Tate Shafarevich group, thereby yielding the claimed three 
different  M-theory vacua.

This note consists of  two sections. In Section \ref{TSphysicsmath} we review some aspects of the 
TS-group and its appearance in M- and F-theory. In particular, we review how the different 
M-theory vacua can be obtained from a circle compactification with discrete choices for the
Wilson line from the unique F-theory compactification, both before and after Higgsing. 
Then, in Section \ref{Threevacua}, we focus on the explicit construction of 
the three different curves supporting the Higgses which lead to the three different M-theory vacua. For 
this analysis we discuss two different resolution phases of the $dP_1$ geometry.

\section{The  Tate-Shafarevich group in M- and F-theory} \label{TSphysicsmath}

In this section we discuss some background material that is necessary for our geometrical analysis in 
section \ref{Threevacua}. We start by elaborating on some mathematical facts about the Tate-
Shafarevich group. Then, we continue with a review and slight clarification
of the field theory formalisms in 6D and 5D developed in \cite{Mayrhofer:2014laa, Mayrhofer:2014haa, Anderson:2014yva, Garcia-Etxebarria:2014qua}.

As already discussed in the introduction, discrete groups naturally arise in F-theory 
compactifications on genus-one fibered Calabi-Yau manifolds $X$ that admit instead of a globally 
well-defined section just an $n$-section. To every such genus-one fibration one can associate its 
corresponding Jacobian fibration $J(X)$, whose fibers are the Jacobian $J(C)$ of the genus one curve 
$C$. The Jacobian $J(C)$ is given by the degree 
zero part of the Picard group Pic$(C)$ which has a distinguished point given by the trivial line bundle. 
Thus,  $J(C)$ is an elliptic curve and $J(X)$ an elliptic fibration.  In addition, the Jacobian $J(C)$ equips 
$C$ with the structure of a homogeneous space over $J(C)$. This additional structure is crucial for the 
notion of the TS-group $\Sha(J(X))$ associated to the Jacobian fibration which comprises 
all genus one fibrations that share the same Jacobian. While the latter statement describes the 
TS-group only as a set, the group structure is related to the Jacobian action on the different genus one 
fibrations $\Sha(J(X))$ and is discussed in further detail in appendix \ref{appendixTS}. 
Since the discriminant locus and $\tau$ function of every element $X$ in $\Sha(J(X))$ are identical to 
those of $J(X)$, their corresponding F-theory compactifications are identical. The TS-group 
$\Sha(J(X))$ is manifest in the corresponding F-theory compactification as its discrete gauge group. In 
particular, we expect $\mathbb{Z}_n\subset \Sha(J(X)$ for a genus-one fibration $X$ with only an 
$n$-section and the discrete gauge group of F-theory to contain a $\mathbb{Z}_n$-factor.

A different point of view on a genus-one fibration admitting an $n$-section is to start with an elliptic 
fibration with $n$ rational sections. Then, one performs a  chain of  complex structure deformations 
that  merge the $n$ sections into a single $n$-section. Technically, if we define the $n$ sections
as the $n$ rational roots of a polynomial, we can understand this deformation as introducing additional 
terms to the polynomial so that its roots are no longer rational, but involve taking roots. In other words,
the sections can only be defined over a certain field extension\footnote{In general, given a field $K$ 
with closure $\bar K$, a divisor $D$ is defined over $K$, if $\sigma(D)=D$ for all elements $\sigma$ in 
the absolute Galois group $G_{\bar K/K}$. Also note that, as we are dealing with a function field 
associated to the two-dimensional base, the notion of roots refers to taking suitable covers of the 
base.}. Upon considering a family of these polynomials parametrized by the base of the elliptic fibration, 
the  $n$ individual roots are not well-defined anymore due to branch cuts around which they are 
exchanged by monodromies.  Only monodromy-invariant combinations are globally well-defined in the 
fibration, giving rise to the $n$-section which is an appropriate union of the original roots.

In contrast to the F-theory picture, one obtains different vacua by compactifying the dual
M-theory on the corresponding elements of the Tate-Shafarevich group. 
These are related to the F-theory vacuum by an additional $S^1$-compactification in which we in 
addition specify a certain flux choice along the compactification
circle.  This flux choice introduces the sought for additional
degree of freedom, that is needed to reproduce the different lower-dimensional M-theory vacua 
starting from a unique F-theory compactifcation before circle compactification, as we explain next.

The following, purely field theoretical analysis of the emergence of the different M-theory vacua has 
been proposed in \cite{Mayrhofer:2014laa, Mayrhofer:2014haa,Anderson:2014yva, Garcia-Etxebarria:2014qua}.  Here, we 
review the aspects essential to our discussion and elucidate some additional important 
insights. Although the following similarly applies to four-dimensional F-theory compactification, we focus 
here on  F-theory in 6D with an M-theory dual in 5D for clarity of our discussion. The crucial point in the following is the fundamental difference between the Higgs effect in 6D and in 5D due to the 
presence of another U(1) gauge field in 5D, that is the Kaluza-Klein U(1).

The starting point is a six-dimensional U(1)$_{6d}$ 
gauge theory that we want to  break by a Higgs field $\Phi$ of charge $q$ to $\mathbb{Z}_q$. As a 
next step, we compactify this theory on a circle $S^1$. The Higgs field enjoys accordingly an expansion 
of the form
\be
\Phi(x,y) = \sum_{n \in \mathbb{Z}} \phi_n(x) e^{2\pi i n y}\, .
\ee
In addition, one can turn on a flux (Wilson line) $\xi= \int_{S^1} A_{6d}$ along the $S^1$. The full 
mass formula for the $n$th KK-mode then reads
\be\label{massformula}
m_n^q = |q \xi + n |\, , 
\ee
where $q$ and $n$ refer to the charges under U(1)$_{6d}$  and U(1)$_{KK}$, the Kaluza-Klein (KK) 
U(1) gauge field, respectively. 
 
It is crucial to note that these gauge groups and their corresponding charges are only well-defined up to unimodular transformations that act on the charge vector $(q,n)$ and the two U(1) fields as
\be
\begin{pmatrix} q & n \end{pmatrix} \longmapsto \begin{pmatrix} q & n \end{pmatrix} \begin{pmatrix} d & -b \\ -c & a \end{pmatrix} \, , \qquad \begin{pmatrix} \text{U(1)}_{6d} \\ \text{U(1)}_{KK} \end{pmatrix} \longmapsto \begin{pmatrix} a & b \\ c & d \end{pmatrix} \begin{pmatrix} \text{U(1)}_{6d} \\ \text{U(1)}_{KK} \end{pmatrix} \, , \quad \begin{pmatrix} a & b \\ c & d \end{pmatrix} \in \text{SL(2,}\mathbb{Z})\,.
\ee
These transformations allow us to identify different 5D KK-towers that arise from different 
choices for the flux $\xi$. Indeed, if one re-defines the five-dimensional U(1) fields according to
\be
\begin{pmatrix}
\text{U(1)}_{6d} \\ \text{U(1)}_{KK}
\end{pmatrix}
\mapsto 
\begin{pmatrix}
1 & -r \\ 0 &1
\end{pmatrix}
\begin{pmatrix}
\text{U(1)}_{6d} \\ \text{U(1)}_{KK}
\end{pmatrix}
\ee
while shifting $\xi \mapsto \xi + r$ at the same time, the mass formula \eqref{massformula} is 
invariant. Thus, one obtains again the same KK-tower. 

Varying the circle-flux $\xi$, we see that we have 
phase transitions at $\xi=\frac{k}{q}$ for $k \in \mathbb{Z}$ since the KK-tower gets shifted, 
according to \eqref{massformula}. For a given flux $\xi$, the mode $\phi_n$ triggering the Higgsing in 
5D, which has to be chosen so that its mass in \eqref{massformula} vanishes, is changed.  Each 
different choice for the 5D Higgs yields a \textit{different} theory after Higgsing. 
Thus, we see that there are $q$ 
inequivalent Higgses (for each different value of $\xi$) and corresponding 
five-dimensional vacua after Higgsing.  These $q$ different vacua precisely 
correspond to the different M-theory compactifications obtained from the different elements
of the TS-group.

As a concrete example, let us focus on the case of interest, which is 6D F-theory compactification with 
one U$(1)_{6d}$ with a Higgs $\Phi$ of charge  $q=3$. This is realized by F-theory compactified on 
$dP_1$-elliptic fibrations \cite{Klevers:2014bqa} and the Higgsed theory has a $\mathbb{Z}_3$ 
discrete group. In 5D, one obtains the following gauge groups depending on the KK-charge of
the 5D Higgs:
\be \label{eq:3dcases}
\begin{cases}
U(1) \times \mathbb{Z}_3, \quad \text{if}\,\, n = 0\,\, \text{mod}\,\, 3 \\
U(1) , \quad \,\, \qquad \text{if}\,\, n \neq 0 \,\,\text{mod}\,\, 3
\end{cases}\, .
\ee
By the above discussion,  we expect $q=3$ M-theory vacua with massless Higgses fixed by 
\eqref{massformula} for $\xi=0,\,\frac{1}{3},\,\frac{2}{3}$, respectively. By \eqref{eq:3dcases} we see that only one of 
which has discrete gauge group $\mathbb{Z}_3$ and 
two of which have no discrete gauge group. Geometrically, this is expected as the TS-group of the 
cubic, which is the geometry obtained by deforming the $dP_1$-elliptic fibration in the Higgsing 
\cite{Klevers:2014bqa}, contains $\mathbb{Z}_3$, see Appendix \ref{appendixTS}. The geometric 
identification of these three Higgs field in terms of holomorphic curves wrapped by M2-branes is 
the subject of Section \ref{Threevacua}.
 
Before turning to this discussion, we would like to briefly mention a dual reformulations of the 
above picture in terms of a St\"uckelberg mechanism 
\cite{Anderson:2014yva, Garcia-Etxebarria:2014qua}. 
We identify an  axion  as the phase of the six-dimensional Higgs field $\Phi = h e^{ic}$. 
In the circle compactification from 6D to 5D, the axion $c$ can acquire a vacuum expectation value
\be
n = \int_{S^1} \langle dc \rangle \label{axionvev}
\ee
along the circle. Clearly, the VEV \eqref{axionvev} signals a non-trivial profile of $c$ along the $S^1$, 
so that the flux \eqref{axionvev} agrees with the KK mode number $n$ of $c$. In this flux background, 
the surviving massless U(1) is a linear combination of the two 5D U(1) fields $U(1)_{6d}$ and 
$U(1)_{KK}$. Geometrically, a non-trivial vacuum expectation value of the axion is exactly expected to 
arise in compactifications involving multi-sections and takes into account the generalized T-duality 
transformation rules that govern the physics in the presence of off-diagonal terms in the metric 
\cite{Anderson:2014yva}.

This picture is dual to the above discussion with a corresponding Wilson line $\xi$ that has to be chosen 
so that the massless Higgs by \eqref{massformula} yields precisely the same linear combination of 
$U(1)_{6d}$ and $U(1)_{KK}$ of the remaining massless U(1) determined by the St\"uckelberg 
mechanism of the 5D axion $c$.

\section{Identifying 5D Higgs Fields for the $\mathbb{Z}_3$ Tate-Shafaverich group} \label{Threevacua}

In this section we construct explicitly the different Higgs fields that lead to the three inequivalent 
five-dimensional M-theory vacua corresponding to the three different elements of the Tate-Shafarevich 
group of the most general cubic.  These Higgs fields are obtained from M2-branes wrapping three 
different holomorphic curves in $I_2$-fibers of the elliptic fibration with fiber in $dP_1$, that we 
identify explicitly. F-theory compactifications both on Calabi-Yau fibrations  with the most general cubic 
and the elliptic curve in $dP_1$ have been studied recently in \cite{Klevers:2014bqa} to which we refer 
for further details.

A genus-one fibration by the most general cubic in $\mathbb{P}^2$ is defined by
the hypersurface 
\be
p_{\text{cub}} = s_1 u^3+s_2 u^2 v+s_3 u v^2+s_4 v^3+s_5 u^2 w+ s_6 u vw +s_7v^2w + s_8 u w^2+ s_9 v w^2 + s_{10} w^3 =0 \, .\label{cubic}
\ee
Here the $s_i$ are polynomials of appropriate degree of the coordinates on $B$ or, more generally,
sections of appropriate line bundles on $B$.
By making specific choices for these line bundles and 
taking general sections, we obtain a genus one fibered Calabi-Yau threefold $X_{p_{\text{cub}}}$.
Clearly, for generic $s_i$ this genus one-fibration admits only a tri-section only.

Our strategy is to start in a six-dimensional F-theory compactification with a U(1) gauge group that we
want to break down to $\mathbb{Z}_3$ by a Higgsing. Such a geometry is provided by the
dP$_1$-elliptic fibration which has been considered in great detail in \cite{Klevers:2014bqa}. 
It is related to the geometry of the cubic by the tuning and resolution of the form
\be \label{eq:HiggsDiagram}
\text{cubic} \overset{s_{10} \rightarrow 0}{\xrightarrow{\hspace*{2cm}} } \text{singular}\,\, \text{dP}_1 \overset{\text{blowing up} \,\, e_1 }{\xrightarrow{\hspace*{2cm}} } \text{resolved}\,\, \text{dP}_1\,,
\ee
which is geometrically a conifold transition. Thus, the singular $dP_1$ model is given \eqref{cubic}
for $s_{10}=0$,
\be
p_{dP_1}^{s}=s_1 u^3+s_2 u^2 v+s_3 u v^2+s_4 v^3+s_5 u^2 w+ s_6 u vw +s_7v^2w + s_8 u w^2+ s_9 v w^2 = 0\, . \label{dP1}
\ee
One observes that it has conifold singularities at co-dimension two in the base at $s_8=s_9=0=u=v$. 
In the following two subsections, we argue that this singularity admits two different 
resolutions related by a flop. These resolutions lead to the same F-theory vacuum, but the 
corresponding M-theory vacua are different as we demonstrate by analysing the charges of the 
corresponding $I_2$-fibers in the respective resolutions.  

The crucial point is that in the first resolution two holomorphic curves are manifest that correspond to 
two of the three wanted 5D Higgs fields, whereas the curve supporting the third Higgs is obscured. In 
contrast, in the second resolution, again two curves are manifest that, however, 
correspond to the \textit{third}  wanted 5D Higgs field, while one of the curves manifest in the first 
resolution is obscured now. Thus, we see that by using different phases of the resolution of the 
singularity at $s_8=s_9=0$ in the $dP_1$-elliptic fibration, we can indeed manifestly see three 
different holomorphic curves to be used as the 5D Higgses yielding the three different M-theory vacua 
from compactification on the three elements of the TS-group of the cubic \eqref{cubic}.

\subsection{Resolving by the toric blow-up}

The obvious resolution of the singularities in  elliptic fibration  \eqref{dP1} is obtained by 
performing the toric blow-up in the ambient space of the elliptic fiber defined by
\be \label{eq:toricBU}
u \longrightarrow e_1 u, \qquad v \longrightarrow e_1 v\,,
\ee
where the coordinate $e_1$ vanishes at the exceptional divisor. This is precisely toric $dP_1$, which 
yields a resolution of the singular fibration with the cubic \eqref{dP1}. The corresponding smooth elliptic 
fibration has been extensively studied in \cite{Klevers:2014bqa} and we just summarize the 
most important results here. First of all the toric data of $dP_1$ is given by
\be
\left(\begin{array}{cc|cc|c|c}
 1 & 0 & 1 &0 & w & U  \\
 -1&1&0&1 & u & U - S\\
 -1&0&1&-1 & e_1 & S \\
 0&-1&0&1 & v & U -S \\
\end{array}\right)\,.
\label{toricdp1}
\ee
Here the first two columns display the points of the toric diagram, written as row  vectors, 
followed by the generators of the Mori cone in the third and fourth column, while the last two colums 
refer to the coordinates and the corresponding divisor classes being assigned to the rays of the 
diagram, respectively. The fan of the toric variety $dP_1$ is obtained by a fine star-triangulation of the
polytope defined by \eqref{toricdp1}, from which one immediately 
obtains the Stanley Reisner ideal as well as the intersection numbers
\be
SR= \{uv, e_1 w \}, \qquad U^2 = 1, \quad S^2 =-1, \quad U\cdot S=0\, .
\ee
The resolved curve takes the form 
\be \label{eq:CYeqdP1}
p_{dP_1}^{r} = s_1 u^3 e_1^2+s_2 u^2 v e_1^2 +s_3 u v^2 e_1^2 +s_4 v^3 e_1^2 +s_5 u^2 w e_1+ s_6 u vw e_1 +s_7v^2w e_1 + s_8 u w^2+ s_9 v w^2 = 0\, .
\ee
Over the locus defined by $s_8=s_9=0$, the fiber degenerates as
\be \label{eq:I2s8s9}
\underbrace{e_1}_{c_1} \underbrace{\left(s_1 u^3 e_1+ s_2 u^2 v e_1+ s_3 u v^2 e_1+ s_4 v^3 e_1+s_5 u^2 w+ s_6 u v w+ s_7 v^2 w \right)}_{c_2} = 0\, ,
\ee
which is a smooth $I_2$-fiber.
Here $c_1$ and $c_2$ denote the two $\mathbb{P}^1$ components of the resulting $I_2$-fiber 
which have classes $S$ and $3U-2S$.

Next, we investigate the sections, starting from the singular model before the blow-up 
\eqref{eq:toricBU}. It has two sections which take the form
\be
S_0 = \left[0:0:1\right], \qquad S_1 = \left[-s_9 : s_8  : \frac{s_4 s_8^3 - s_3 s_8^2 s_9 + s_2 s_8 s_9^2 - s_1 s_9^3 }{-s_7 s_8^2 + 
 s_6 s_8 s_9 - s_5 s_9^2} \right]\, \label{singularsection}
\ee
in terms of the projective coordinates $[u:v:w]$ on $\mathbb{P}^2$.
While the first section is toric and taken to define the zero-point $O$ on the elliptic curve, the second 
section is found by constructing the tangent line
\be
t_{O} : s_8 u + s_9 v =0 
\ee
to the origin $O$ which has to intersect the cubic in precisely one third point, provided 
$s_8 \neq 0$ or $ s_9 \neq 0 $. This third intersection point is rational and defines the section $S_1$. 
In contrast, at the singular locus $s_8=s_9=0$, the curve becomes singular precisely at $O$ which 
implies that any line through $O$ is tangent and therefore $S_1$ degenerates to the whole singular 
curve. The two curves $c_1$ and $c_2$ support the Higgs field in question.

Passing to the resolved geometry, the two sections \eqref{singularsection} lift to
\be
\tilde S_0 = [s_9:-s_8:1:0]\, , \qquad  \tilde S_1 = \left[-s_9 : s_8  : s_4 s_8^3 - s_3 s_8^2 s_9 + s_2 s_8 s_9^2 - s_1 s_9^3 : -s_7 s_8^2 + 
 s_6 s_8 s_9 - s_5 s_9^2 \right]\, .
\ee
Note that these expressions are only valid away from the locus $s_8=s_9=0$. In fact, $\tilde S_0$ as 
well as $\tilde S_1$ are rational sections that wrap the components $c_1$ and $c_2$ completely, 
respectively. 
We note that the classes of the divisors associated to the sections are given by
\be \label{eq:classS0S1}
[\tilde S_0] \cong S, \qquad [\tilde S_1] \cong U -2S\, ,
\ee
where, by abuse of notation, we denoted the divisors $S$ and $U$, defined in \eqref{toricdp1}, on the 
ambient space by the same symbols as their intersections with the Calabi-Yau hypersurface 
\eqref{eq:CYeqdP1}.
While the class of the zero section is obvious, the class of the second section is obtained by noting that the tangent line defines a tri-section. Subtracting from its class two times the class of the zero section\footnote{Recall that the tangent line was defined in the singular geometry, where the class of $u$ reads $U$ instead of $U-S$.}, one is left with $U-2S$. 

Knowing the classes of the sections as well as those of the fiber components $c_1, c_2$, one can 
evaluate the corresponding charges w.r.t.~the U$(1)_{6d}$ and the KK U(1)-field U$(1)_{KK}$.
Recalling  that the charge under U$(1)_{6d}$ is computed from the intersection of the Shioada map
for $\tilde{S}_0$, that is $\tilde{S}_1-\tilde{S}_0$, with the respective curve $c_i$ while the 
KK-charge is given by the intersection number with the zero section $\tilde{S}_0$, we obtain the 
charges summarized in Table \ref{intersectiontabledP1}. One observes that 
the shrinking of $c_1$ and $c_2$ gives massless states with charges $(-3,-1)$ and $(3,2)$, 
respectively. According to \eqref{massformula}, these states are massless for the associated  (inverse) 
choices of flux $\xi_1 = -\frac{1}{3}\text{ mod } 1$ and  $\xi_2 = -\frac{2}{3} \text{ mod } 1$. 
By \eqref{eq:3dcases} the remaining gauge group in 5D after Higgsing is U(1).
\begin{table} 
\begin{center}
\begin{tabular}{|c|c|c|} \hline
& $c_1$ & $c_2$ \\\hline
$\tilde S_0$ \rule{0pt}{1Em}& $-1$ & $2$ \\\hline
$\tilde S_1$\rule{0pt}{1Em} & $2$ & $-1$ \\\hline
$\tilde S_1\, - \, \tilde S_0$ \rule{0pt}{1Em}& $3$ & $-3$ \\\hline
\end{tabular}
\caption{Intersection numbers of the sections $\tilde{S}_0$, $\tilde{S}_1$ with the curves in the
$I_2$-fiber at $s_8=s_9=0$ and corresponding U$(1)_{6d}$ charges in the toric resolution of the dP$_1$-model.} \label{intersectiontabledP1}
\end{center}
\end{table}

Geometrically, we have to shrink the respective curves $c_1$ and $c_2$ and deform the arising 
singularity, cf.~the inverse process of \eqref{eq:HiggsDiagram}.  While the shrinking of the first 
component $c_1$ is performed by taking the blow-down $e_1\rightarrow 1$, the shrinking of the 
second component $c_2$ is more elaborate. Instead of describing this directly, we choose the following 
short-cut. Observe that the charges in Table \ref{intersectiontabledP1} are symmetric in the curves 
$c_1$ and $c_2$ as well as in $\tilde{S}_0$ and $\tilde{S}_1$. Therefore, we can also change the 
roles of $\tilde{S}_0$ and $\tilde{S}_1$, i.e.~we take $\tilde{S}_1$ to be the zero 
section\footnote{As shown in \cite{Grimm:2015zea}, such a  change of the choice of zero section is 
always possible in an anomaly-free theory, which is the case for the theory at hand, 
cf.~\cite{Klevers:2014bqa}.}. Then, we obtain the 
second vacuum again by blowing down $e_1 \rightarrow 1$. Geometrically, this gives the same cubic 
after Higgsing which corresponds to switching on a non-zero value for $s_{10}$, 
However, this is precisely what we expect. The two non-trivial elements of the Tate Shafarevich group 
are given by the same cubic that only differs through inverse actions of the Jacobians, cf.~Appendix 
\ref{appendixTS}. 

It remains to show how to obtain the massless mode of the Higgs that yields the third 
vacuum which must correspond to the Jacobian. 
The relevant Higgs field has to have charge $(3,0)$ with $\xi=0$, so that the gauge group
is $\mathbb{Z}_3$, according to \eqref{eq:3dcases}. However, the associated curve is not manifest in 
the Calabi-Yau manifold \eqref{eq:CYeqdP1}. Nevertheless, one can argue for its existence by 
studying the enumerative geometry of the $dP_1$-fibration, which can be accomplished evaluating 
the non-perturbative part of the prepotential of the topological string, 
\be
F^0_{\text{non-pert}} = \sum_{\beta} N_{\beta}^0 Q^\beta\,,
\ee
analogously to the analysis performed in \cite{Klemm:1996hh}. Here, $N_{\beta}^0$ denote the 
(genus zero) Gromov-Witten invariants of the classes $\beta$. 

To this end, we construct explicitly
all Calabi-Yau manifolds \eqref{eq:CYeqdP1} with base $B=\mathbb{P}^2$ following the algorithm in 
Appendix G of \cite{Cvetic:2013uta} and compute all 
Gromov-Witten invariants up to a suitable degree. We find that in all these cases the
invariant $N_{[c_1+T^2]} = [s_8]\cdot [s_9]$, i.e.~agrees precisely with the number of $I_2$-fibers
at $s_8=s_9=0$. This is a strong indication that there exists a third holomorphic curve of genus zero
inside the $I_2$-fiber \eqref{eq:I2s8s9} at $s_8=s_9=0$, that is harder to manifestly see in geometry. 
In fact, using the classes \eqref{eq:classS0S1} we check that this curve has precisely the right charges, 
$(q,n)=(3,0)$, that are required for obtaining the third geometry according to \eqref{eq:3dcases}. 

Next we consider a different resolution which makes the curves in the classe $[c_1+T^2]$
directly manifest in the geometry.

\subsection{Resolving by a complete intersection resolution}

Let us begin by noting that in the toric blow-up \eqref{eq:toricBU}, the zero section is ill-behaved, 
i.e.~wraps a fiber component, precisely at the $I_2$-locus of interest at $s_8=s_9=0$. Therefore,
the two manifest curves $c_1$ and $c_2$ in \eqref{eq:I2s8s9} have to have a non-trivial KK-charge by 
construction. One way to manifestly see the third curve at $s_8=s_9=0$ with KK-charge zero is
to consider a different phase of the resolution of the $dP_1$-elliptic fibration \eqref{dP1} where the 
zero section remains \textit{holomorphic}. In the following, we construct this resolution explicitly and 
show the existence of the desired curve with KK-charge zero. We note that such a resolution has been 
considered in a different context in \cite{Braun:2011zm,Borchmann:2013hta}.

To get started, we rewrite the singular $dP_1$ geometry \eqref{dP1} in the form of a 
determinental variety,
\be \label{eq:detVariety}
p_{dP_1}^{s} = u P_1 - v P_2\,.
\ee
Here we have defined the two summands as
\be
P_1 = s_1 u^2+s_2 u v+s_3  v^2+s_5 u w+ s_6  vw  + s_8  w^2\, , \quad
P_2 = -\left( s_4 v^2 + s_7 v w+ s_9 w^2\right) \, . \label{P1constraint} 
\ee
For a variety of the form \eqref{eq:detVariety}, we can perform a small resolution. This means we introduce the coordinates $\lambda_1, \lambda_2$ parameterizing a $\mathbb{P}^1$ and impose
\be
u \lambda_1 = \lambda_2 P_2\,. \label{nonstandardblowup}
\ee
The proper transform of the $dP_1$ geometry \eqref{dP1} is accordingly given by
\be
v \lambda_1 = \lambda_2 P_1\,. \label{propertransform}
\ee
In contrast to the toric blow-up in \eqref{eq:toricBU} that does not change the dimension of the 
ambient space, the two equations \eqref{nonstandardblowup} and \eqref{propertransform} define the 
resolved fiber as a complete intersection in a three-dimensional ambient space that can be torically 
characterized as
\be
\left(\begin{array}{ccc|cc|c|c}
  -1 & -1 & -1 &0&1 & u & L \\
 -1&0&0&1&-1 & \lambda_2 & H \\
 0&0&1&0&1& v & L \\
 0&1&0&0&1 & w & L \\
 1&0&0&1&0&\lambda_1 & H+L
\end{array}\right)\, .
\label{toricp1p2}
\ee
Again, the first three columns specify the points of the toric diagram, written as row vectors, 
followed by the two generators of the Mori cone. The last two columns display the coordinates and the 
divisor classes which are associated to the rays, respectively.  From the toric data one obtains 
immediately the Stanley Reisner ideal as well as the triple intersection numbers as
\be
SR=\{\lambda_1 \lambda_2, u v w \}, \qquad L^3=0, \qquad L^2 \cdot H =1, \qquad L\cdot H^2=-1, \qquad H^3=1\, . \label{intersectionnumbers}
\ee
The evaluation of the irreducible components of the $I_2$-fiber is slighly more involved in this case and 
is best performed using a primary decomposition. At the locus $s_8 = s_9 = 0$ we find that the two 
rational components of the $I_2$ fiber are given by
\begin{eqnarray}
c_+ &=& V\left(u,v\right)\,, \\
c_- &=& V\left( s_4 v^2 \lambda_2+s_7v w \lambda_2+u \lambda_1,\, s_1 u^2 \lambda_2+s_2 u v \lambda_2+s_3 v^2 \lambda_2+s_5 u w \lambda_2+s_6 v w  \lambda_2-v \lambda_1, \right. \nn \\ &&
\left. \,\,\,\,\,\,\,\,  s_1 u^3+s_2 u^2 v+s_3 u v^2+s_4 v^3+s_5 u^2 w+s_6 u v w+s_7 v^2 w, \,
   s_1 s_4 u v \lambda_2^2 +s_1 s_7 u w \lambda_2^2\right. \nn \\ &&\,\,\,\,\,\,\,\, \left. +s_4 s_5 v w \lambda_2^2+s_5 s_7 w^2 \lambda_2^2-s_2 u \lambda_1 \lambda_2-s_3 v \lambda_1 \lambda_2-s_6 w \lambda_1 \lambda_2+ \lambda_1^2 \right)\, .   \nn\label{c2ideal}
\end{eqnarray}
Here, $V(I)$ denotes the zero set of the ideal $I$ and $(p_1,\cdots,p_k)$ denotes the ideal generated 
bythe polynomials $p_1,\, \ldots,\,p_k$.
Applying again prime ideal techniques, one determines their respective homology classes of the curves $c_+$ and $c_-$ as
\be \label{eq:I2curvesSecondRes}
[c_+] = L^2, \qquad [c_-] = (2L+H)(2L+H)-L^2 \, .
\ee

Next, we turn to the analysis of the sections \eqref{singularsection}. Outside $s_8=s_9=0$ the 
simultaneous vanishing of $u$ and $v$ implies that $\lambda_2=0$ as well. Thus, we identify the 
coordinates $[u:v:w:\lambda_1:\lambda_2]$ of the zero section and its homology class as
\be \label{eq:S0'}
  S_0' = [0:0:1:1:0] \, , \qquad [ S_0'] = H\,.
\ee
In contrast to the toric blow-up, $S_0'$ is also well-defined at the singular locus and defines, therefore.
a holomorphic zero section. In order to find the second section, we note that away from the singular 
locus there is a unique solution for $\lambda_1$ and $\lambda_2$ if one plugs the coordinates of the 
section $S_1$ given in \eqref{singularsection} into \eqref{nonstandardblowup} and 
\eqref{propertransform}. Its coordinates are given by
\begin{eqnarray}
S_1'\! \!&\!\!=\!\!&\!\!\! [s_9:-s_8:  \frac{-s_4 s_8^3 + 
 s_9 (s_3 s_8^2 + s_9 (-s_2 s_8 + s_1 s_9))}{s_7 s_8^2 + s_9 (-s_6 s_8 + s_5 s_9)} :\big(-s_4 s_6 s_7 s_8^5 + s_3 s_7^2 s_8^5 + s_4^2 s_8^6+ s_4 s_6^2 s_8^4 s_9 \nn \\ \!\!&+\!\!&\!\!\!
 s_4 s_5 s_7 s_8^4 s_9 - s_3 s_6 s_7 s_8^4 s_9 - s_2 s_7^2 s_8^4 s_9- 
 2 s_3 s_4 s_8^5 s_9 - 2 s_4 s_5 s_6 s_8^3 s_9^2 + s_3 s_5 s_7 s_8^3 s_9^2+ 
 s_2 s_6 s_7 s_8^3 s_9^2\nn \\\!\!&+\!\!&\!\!\!   s_1 s_7^2 s_8^3 s_9^2 + s_3^2 s_8^4 s_9^2+ 
 2 s_2 s_4 s_8^4 s_9^2 + s_4 s_5^2 s_8^2 s_9^3 - s_2 s_5 s_7 s_8^2 s_9^3- 
 s_1 s_6 s_7 s_8^2 s_9^3 - 2 s_2 s_3 s_8^3 s_9^3+ s_1^2 s_9^6\nn \\\!\!&-\!\!&\!\! \!  2 s_1 s_4 s_8^3 s_9^3+ 
 s_1 s_5 s_7 s_8 s_9^4 + s_2^2 s_8^2 s_9^4 + 2 s_1 s_3 s_8^2 s_9^4 - 
 2 s_1 s_2 s_8 s_9^5 \big)\!:\!  (s_7 s_8^2 + s_9 (  s_5 s_9-s_6 s_8))^2]. \label{resolvedsection}
\end{eqnarray}
A slightly alternative way to obtain this section is to compute the intersection of the hyperplane given by
\be
s_8 u + s_9 v =0
\ee
with the complete intersection manifold specified by \eqref{nonstandardblowup} and 
\eqref{propertransform}.
Analogous to the case of the toric blow-up, its class is computed as
\be \label{eq:S1'}
[S_1'] = [u]-2[\tilde S_0] = L-2H\, .
\ee

Finally, we compute the one finds the charges to the two rational curves $c_+$, $c_-$ in the  
$I_2$-fiber at $s_8=s_9=0$ in this resolution. Here, we use their homology classes in  
\eqref{eq:I2curvesSecondRes} and the homology classes of the zero and rational sections in 
\eqref{eq:S0'} and  \eqref{eq:S1'} to obtain the charges summarized in Table 
\ref{intersectiontabledP1diff}. 
\begin{table} 
\begin{center}
\begin{tabular}{|c|c|c|} \hline
& $c_+$ & $c_-$ \\\hline
$ S_0'$\rule{0pt}{1Em} & $1$ & $0$ \\\hline
$S_1'$\rule{0pt}{1Em} & $-2$ & $3$ \\\hline
$S_1'\, - \,  S_0'$\rule{0pt}{1Em} & $-3$ & $3$ \\\hline
\end{tabular}
\caption{Intersection numbers of the sections $S_0'$, $S_1'$ with the curves in the
$I_2$-fiber at $s_8=s_9=0$ and corresponding U$(1)_{6d}$ charges in the complete intersection 
resolution of the dP$_1$-model.} \label{intersectiontabledP1diff}
\end{center}
\end{table}
In particular, we observe that one obtains a mode with KK-charge zero by shrinking the component 
$c_-$. It supports precisely the third Higgs necessary to obtain the third 5D M-theory vacuum with 
$\text{U}(1)\times\mathbb{Z}_3$ gauge group in \eqref{eq:3dcases}. We emphasize that the curve 
$c_-$ has precisely the same charges as the curve in the class $[c_1+T^2]$ for whose existence we 
argued in the last subsection by computing the corresponding Gromov-Witten invariants.  In analogy to 
the analysis performed in 
\cite{Mayrhofer:2014haa, Mayrhofer:2014laa}, it is expected that the corresponding geometry after 
the deformation is given by the Jacobian of the most general cubic \eqref{cubic}.

\section{Conclusion and further directions}

In this note we have considered the physics of F-theory compactifications with $\mathbb{Z}_3$ 
discrete gauge group using M-/F-theory duality. 
From a mathematical perspective this discrete gauge group arises from compactifcation on
a genus-one fibered Calabi-Yau manifold with Tate-Shafarevich group $\mathbb{Z}_3$. While 
the effective physics of the F-theory compactification is the same, regardless of which element of 
$\Sha(J(X))$ is chosen, the compactification of M-theory to 5D on these three Calabi-Yau manifolds 
gives rise to three physically different vacua. We understand the emergence of these three different 
M-theory vacua by starting with an F-theory compactification with a U(1) and matter with charge three
obtained by compactification on the $dP_1$-elliptic fibration considered in \cite{Klevers:2014bqa}. 
In 6D, Higgsing the U(1) by the charged matter  yields a $\mathbb{Z}_3$ discrete gauge group. 
In 5D, the three different  vacua arise by three different Higgs effects. When performing a fluxed circle 
compactification of the  F-theory \textit{with} the U(1) gauge symmetry, there are three different flux 
choices (U(1) Wilson lines) that yield three different possible massless Higgs fields in 5D. The 
KK-labels of these Higgs fields are $n=0,1,2$, respectively, that agree with their charge with respect 
to the 5D KK U(1) field. 
The three different vacua of M-theory are then realized by these  three different Higgsings: for the two 
Higgsing by the fields with $n=1,2$, we obtain a 5D theory with a remaining U(1) gauge group, while we 
obtain a U(1)$\times\mathbb{Z}_3$ gauge group for the third Higgsing by the field with $n=0$. 
The crucial advancement of our analysis is the explicit construction of the \textit{three} different vanishing
rational curves in a single $I_2$-fiber of the $dP_1$-elliptic fibration that yield the desired Higgs fields 
from wrapping M2-brane states. 

We collect first evidence for the existence of three rational curves in a single $I_2$-fiber in the 
$dP_1$-elliptic  fibration, by computation of the Gromov-Witten invariants in the three classes in
which these curves are expected to be.
In order to identify these three curves explicitly and the corresponding three different vacua, it was 
crucial to consider two different phases of the resolution  of the $dP_1$-elliptic fibration. The first one 
of these 
is the standard toric blow-up in $dP_1$. Due to the fact  that in this resolution the zero section is 
rational and wraps a fiber component precisely at the codimension two locus in the base, where the 
$I_2$-fiber is located, only the curves supporting the Higgs fields with KK-charges $n=1,2$ are 
manifest. In contrast, the resolution in the second phase was performed using a small resolution, 
resulting 
in an elliptic fibration with elliptic fiber defined as a complete intersection in a three-dimensional ambient 
space. The key point is that in this fibration the zero section is holomorphic. Thus,  the zero section is
well-behaved everywhere on the base, in particular at the aforementioned $I_2$-locus. Consequently, 
there exists a curve at the $I_2$-locus with KK-charge $n=0$, yielding the third Higgs, that was 
invisible in the first phase described by the toric blow-up. The geometry 
that arises by performing the conifold transition of this curve in the $I_2$-locus is expected to be 
the Jacobian of the genus-one fibration, i.e.~the trivial element in the TS-group. The other two 
geometries that arise from performing the conifold transition for the 
two curves with KK-charge $n=1,2$ in the toric blow-up, respectively, give both rise to the  most 
general cubic corresponding to two M-theory vacua with no discrete gauge symmetry. 
However, their KK-towers are inequivalent. Mathematically, 
the only way to distinguish these two geometries is given by an additional piece of data given by the 
Jacobian action, which is in fact inverse for these two geometries.

While the analysis of the three different vacua has been successfully completed which was the main aim 
of this note, there are is a list of interesting questions that demand further investigation and will be 
addressed in future works \cite{workinprogress}:
\begin{enumerate}
\item What is the interpretation of the charge content in terms of intersection theory? It seems that is 
hard to achieve the charge content displayed in table \ref{intersectiontabledP1diff} using the methods 
presented in \cite{Morrison:2012ei} for rational curves with self-intersection $-1$. 
\item What is the physical interpretation of the Jacobian action on the genus-one geometries? This 
action is crucial to differentiate the two elements of the Tate Shafarevich group which are geometrically 
identical.
\item How does the group structure carried by $\Sha(J(X))$ enter the physical discussion? It is 
tempting to conjecture that it is identical to the addition of the discrete Wilson line that is put on the  
circle used to compactify from F- to M-theory.
\item As the third M-theory vacuum corresponding to the mode with charges $(3,0)$ exhibits a 
$\mathbb{Z}_3$-symmetry already in five dimensions, it is expected that the corresponding geometry 
has non-trivial torsional contributions to its homology lattice by analogy with the discussion
in \cite{Mayrhofer:2014laa} on the $\mathbb{Z}_2$ case. It would be interesting to investigate the 
r\^ole of torsion in this example as well. 
\end{enumerate}

\subsubsection*{Acknowledgments}
We would like to thank Antonella Grassi, Thomas Grimm and Albrecht Klemm for helpful discussions and 
comments. MP would like to thank the Bethe Center for Theoretical Physics for hospitality. This work is 
supported by the DOE grant DE-SC0007901 (MC, HP, MP), the NSF String Vacuum Project 
Grant No. NSF PHY05-51164 (HP), the Fay R. and Eugene L. Langberg Endowed 
Chair (MC) and the Slovenian Research Agency (ARRS) (MC).  RD acknowledges partial support by NSF 
grant DMS 1304962.

\begin{appendix}
\section{The Tate-Shafarevich group} \label{appendixTS}
Let $f_0: X_0 \longrightarrow B$ be an elliptically fibred Calabi-Yau fibered over a base $B$. We define the Tate-Shafarevich group as a set as
\begin{eqnarray}
\Sha(X_0) &=& \left\{ \text{equivalence classes of pairs $(f,i)$, where $f: X \rightarrow B$ is a genus one fibration and} \right. \nn \\ &&  \left. i: X_0 \rightarrow J(X/B)\,\, \text{is an isomorphism.} \right\}
\end{eqnarray}
Here we have denoted by $J(X/B)$ the relative Jacobian of $X$ over $B$.
That means that any element $X$ of $\Sha(X_0)$ is a homogeneous space over $X_0$, i.e.~we have a map
\be
a_i: X_0 \times X \longrightarrow X \label{actionhomogeneous}
\ee
such that the following properties hold
\begin{eqnarray}
& a_i(0,x_i) = x_i \qquad \forall x_i \in X\,, &\\ \label{rule1}
 &a_i(x_0,a_i(\tilde x_0,x_i)) = a_i(x_0 + \tilde x_0,x_i)\,,& \\  \label{rule2}
&\forall x_1, \tilde x_1 \in X\,\, \exists! \,\, x_0 \in X_0 \,\, \text{so that} \,\, a_i(x_0,x_1)=\tilde x_1\,.& \label{rule3}
\end{eqnarray}
Therefore, by an element $X_i \in \Sha(X_0)$ we mean the triple
\be
\textbf{X}_i = \left(X_i, f_i, a_i \right)\,.
\ee
We define the addition of two elements $\textbf{X}_i, \textbf{X}_j \in \Sha(X_0)$ as follows. As a manifold, one has
\be
X_{i+j} = \left( X_i \times_B X_j\right)/\sim \,\, =\,\, \left\{(x_i,x_j) | f_i(x_i) =f_j(x_j) \right\}/\sim \, .
\ee
Here, the equivalence relation on two points $(x_i, x_j),\,  (y_i, y_j) \in  X_i \times_B X_j$ is defined 
by
\begin{eqnarray}
(x_i, x_j) \sim (y_i, y_j) \,\,\,\,\,\Leftrightarrow &\exists x_0 \in X_0\, \,  \text{s.t.} \, \, \left(a_i (x_0, x_i), a_j^{-1}(x_0, x_j) \right) = (y_i, y_j)\,,  \label{equivrelation}
\end{eqnarray}
where we have defined $a_i^{-1}(x_0,x_i) := a_i(-x_0,x_i)$.
There is also an action of $X_0$ on $X_{i+j}$ given by
\ba
a_{i+j} : X_0 \times X_{i+j} &\longrightarrow& X_{i+j} \nn \\
a_{i+j} \left(x_0, (x_i,x_j)\right) &\mapsto & \left(a_i (x_0,x_i), x_j\right) 
\ea
Finally, the inverse of an element $\textbf{X}_i$ specified by the triple
\be
\left(X_i, f_i, a_i \right)
\ee
is given by
\be
\textbf{X}_{-i} = \left(X_i, f_i, a_i^{-1} \right)\, . \label{inverse}
\ee

We end by showing that this is really the inverse, i.e. 
\be
\textbf{X}_i + \textbf{X}_{-i} = \textbf{X}_0
\ee
For this, we observe that the map defined by
\be
\varphi: X_i \times_B X_i \,\, \longrightarrow X_0, \qquad \varphi (x_i, \tilde x_i) = x_i - \tilde x_i
\ee
becomes an isomorphism after we divide the LHS by the following 
identification
\be
(x_i, \tilde x_i) \sim (y_j, \tilde y_j) \Leftrightarrow \exists x_0 \in X_0 \,\, \text{s.t.}\,\, (a_i(x_0, x_i), a_i( x_0, \tilde x_i)) =  (y_i, \tilde y_i)\,.
\ee
However this is precisely the equivalence relation on $X_i \times_B X_i^{-1}$ as given in \eqref{equivrelation} if one takes into account that $a_i$ acts as its inverse on the second factor.

Applying the above results to the case of the cubic, which has an order three element in the TS-group, 
we see that it consists of only two different geometries: the cubic and its Jacobian. However, we obtain 
three  elements since the cubic can be endowed with two different actions of the Jacobian, namely a 
chosen action and its inverse. The two cubics obtained this way are their respective inverses in the TS-group.

\end{appendix}


\addcontentsline{toc}{section}{References}

\begin{thebibliography}{10}

\bibitem{Ibanez:1991hv}
  L.~E.~Ibanez and G.~G.~Ross,
  ``Discrete gauge symmetry anomalies,''
  Phys.\ Lett.\ B {\bf 260} (1991) 291.
  
\bibitem{Ibanez:1991pr}
  L.~E.~Ibanez and G.~G.~Ross,
  ``Discrete gauge symmetries and the origin of baryon and lepton number conservation in supersymmetric versions of the standard model,''
  Nucl.\ Phys.\ B {\bf 368} (1992) 3.
  
\bibitem{Dreiner:2005rd}
  H.~K.~Dreiner, C.~Luhn and M.~Thormeier,
  ``What is the discrete gauge symmetry of the MSSM?,''
  Phys.\ Rev.\ D {\bf 73} (2006) 075007
  [hep-ph/0512163].

\bibitem{Banks:2010zn}
  T.~Banks and N.~Seiberg,
  ``Symmetries and Strings in Field Theory and Gravity,''
  Phys.\ Rev.\ D {\bf 83} (2011) 084019
  [arXiv:1011.5120 [hep-th]].

\bibitem{de Boer:2001px}
  J.~de Boer, R.~Dijkgraaf, K.~Hori, A.~Keurentjes, J.~Morgan, D.~R.~Morrison and S.~Sethi,
  ``Triples, fluxes, and strings,''
  Adv.\ Theor.\ Math.\ Phys.\  {\bf 4} (2002) 995
  [hep-th/0103170].
  
\bibitem{Braun:2014oya}
  V.~Braun and D.~R.~Morrison,
  ``F-theory on Genus-One Fibrations,''
  JHEP {\bf 1408} (2014) 132
  [arXiv:1401.7844 [hep-th]].

\bibitem{Morrison:2014era}
  D.~R.~Morrison and W.~Taylor,
  ``Sections, multisections, and U(1) fields in F-theory,''
  arXiv:1404.1527 [hep-th].
  
\bibitem{Anderson:2014yva}
  L.~B.~Anderson, I.~García-Etxebarria, T.~W.~Grimm and J.~Keitel,
  ``Physics of F-theory compactifications without section,''
  JHEP {\bf 1412} (2014) 156
  [arXiv:1406.5180 [hep-th]].
  
\bibitem{Klevers:2014bqa}
  D.~Klevers, D.~K.~Mayorga Pena, P.~K.~Oehlmann, H.~Piragua and J.~Reuter,
  ``F-Theory on all Toric Hypersurface Fibrations and its Higgs Branches,''
  JHEP {\bf 1501} (2015) 142
  [arXiv:1408.4808 [hep-th]].  
  
\bibitem{Garcia-Etxebarria:2014qua}
  I.~García-Etxebarria, T.~W.~Grimm and J.~Keitel,
  ``Yukawas and discrete symmetries in F-theory compactifications without section,''
  JHEP {\bf 1411} (2014) 125
  [arXiv:1408.6448 [hep-th]].

    
\bibitem{Mayrhofer:2014haa}
  C.~Mayrhofer, E.~Palti, O.~Till and T.~Weigand,
  ``Discrete Gauge Symmetries by Higgsing in four-dimensional F-Theory Compactifications,''
  JHEP {\bf 1412} (2014) 068
  [arXiv:1408.6831 [hep-th]].

\bibitem{Mayrhofer:2014laa}
  C.~Mayrhofer, E.~Palti, O.~Till and T.~Weigand,
  ``On Discrete Symmetries and Torsion Homology in F-Theory,''
  arXiv:1410.7814 [hep-th].

\bibitem{Braun:2014qka}
  V.~Braun, T.~W.~Grimm and J.~Keitel,
  ``Complete Intersection Fibers in F-Theory,''
  arXiv:1411.2615 [hep-th].


\bibitem{Borchmann:2013jwa} 
  J.~Borchmann, C.~Mayrhofer, E.~Palti and T.~Weigand,
  ``Elliptic fibrations for $SU(5)\times U(1)\times U(1)$ F-theory vacua,''
  Phys.\ Rev.\ D {\bf 88}, no. 4, 046005 (2013)
  [arXiv:1303.5054 [hep-th]].

\bibitem{Cvetic:2013nia}
  M.~Cveti\v c, D.~Klevers and H.~Piragua,
  ``F-Theory Compactifications with Multiple U(1)-Factors: Constructing Elliptic Fibrations with Rational Sections,''
  JHEP {\bf 1306} (2013) 067
  [arXiv:1303.6970 [hep-th]].

\bibitem{Cvetic:2013jta} 
  M.~Cveti\v c, D.~Klevers and H.~Piragua,
  ``F-Theory Compactifications with Multiple U(1)-Factors: Addendum,''
  JHEP {\bf 1312}, 056 (2013)
  [arXiv:1307.6425 [hep-th]].

\bibitem{Cvetic:2013qsa} 
  M.~Cveti\v c, D.~Klevers, H.~Piragua and P.~Song,
  ``Elliptic fibrations with rank three Mordell-Weil group: F-theory with U(1) x U(1) x U(1) gauge symmetry,''
  JHEP {\bf 1403}, 021 (2014)
  [arXiv:1310.0463 [hep-th], arXiv:1310.0463].


\bibitem{Klemm:1996hh}
  A.~Klemm, P.~Mayr and C.~Vafa,
  ``BPS states of exceptional noncritical strings,''
  Nucl.\ Phys.\ Proc.\ Suppl.\  {\bf 58} (1997) 177
  [hep-th/9607139].
\bibitem{Borchmann:2013hta}
  J.~Borchmann, C.~Mayrhofer, E.~Palti and T.~Weigand,
  ``SU(5) Tops with Multiple U(1)s in F-theory,''
  Nucl.\ Phys.\ B {\bf 882} (2014) 1
  [arXiv:1307.2902 [hep-th]].
  
\bibitem{Morrison:2012ei}
  D.~R.~Morrison and D.~S.~Park,
  ``F-Theory and the Mordell-Weil Group of Elliptically-Fibered Calabi-Yau Threefolds,''
  JHEP {\bf 1210} (2012) 128
  [arXiv:1208.2695 [hep-th]].  
  
\bibitem{Witten:1996qb}
  E.~Witten,
  ``Phase transitions in M theory and F theory,''
  Nucl.\ Phys.\ B {\bf 471} (1996) 195
  [hep-th/9603150]. 
  
\bibitem{Grimm:2015zea}
  T.~W.~Grimm and A.~Kapfer,
  ``Anomaly Cancelation in Field Theory and F-theory on a Circle,''
  arXiv:1502.05398 [hep-th].
  
\bibitem{Cvetic:2013uta} 
  M.~Cveti\v c, A.~Grassi, D.~Klevers and H.~Piragua,
  ``Chiral Four-Dimensional F-Theory Compactifications With SU(5) and Multiple U(1)-Factors,''
  JHEP {\bf 1404}, 010 (2014)
  [arXiv:1306.3987 [hep-th]].
  
  
\bibitem{Braun:2011zm} 
  A.~P.~Braun, A.~Collinucci and R.~Valandro,
  ``G-flux in F-theory and algebraic cycles,''
  Nucl.\ Phys.\ B {\bf 856}, 129 (2012)
  [arXiv:1107.5337 [hep-th]].
  
\bibitem{workinprogress}
M.~Cveti\v{c},  R.~Donagi, D.~Klevers, H.~Piragua and M.~Poretschkin,
\textit{work in progress.} 
  
  
  
\end{thebibliography}
\end{document}